\documentclass[12pt]{article}
\catcode`\@=11
\@addtoreset{equation}{section}
\catcode`\@=12

\usepackage{amsbsy}
\newcommand{\B}{\mathbf B}
\newcommand{\A}{\mathbf A}
\newcommand{\J}{\mathbf J}
\newcommand{\abf}{\mathbf a}
\newcommand{\bbf}{\mathbf b}
\newcommand{\cbf}{\mathbf c}
\newcommand{\vbf}{\mathbf v}
\newcommand{\At}{\tilde \mathbf A}
\newcommand{\X}{\boldsymbol{\times}}
\newcommand{\x}{\bf x}
\newcommand{\rv}{\bf r}
\newcommand{\kv}{\mathbf k}
\newcommand{\D}{\mathbf D}
\newcommand{\W}{\mathbf W}
\newcommand{\del}{\boldsymbol{\nabla}}
\newcommand{\ru}{\hat \mathbf r}
\newcommand{\nn}{\hat \mathbf n}
\newcommand{\thetau}{\hat{\boldsymbol{\theta}}}
\newcommand{\phiu}{\hat{\boldsymbol{\phi}}}

\newcommand{\bea}{\begin{eqnarray}}
\newcommand{\eea}{\end{eqnarray}}
\title{{\small\hfill IMSc/2008/09/14}\\
\textbf{Accounting for monopole configurations in \\
       Yang-Mills theory in three Euclidean dimensions }}
\author{Indrajit Mitra$^{a}$\footnote{E-mail: imitra@iitk.ac.in} ~and
H. S. Sharatchandra$^{b}$\footnote{E-mail: sharat@imsc.res.in} \\\\
$^a$ Department of Physics, Indian Institute of Technology,\\ Kanpur 208016, India\\
$^b$ The Institute of Mathematical Sciences, C.I.T. Campus, Taramani P.O.,\\
Chennai 600113, India}

\date{}

\begin{document}
\maketitle
\begin{abstract}
A gauge transformation provided by the three eigenfunctions of 
$\B^a(x) \cdot \B^b(x)$ (where $\B^a(x)$, with $a=1,2,3$, are the non-Abelian magnetic fields) 
exposes the topological configurations of the Yang-Mills fields. In particular, it gives
Dirac monopoles interacting with `photons' and massless charged vector bosons. A
magnetic dipole field at each monopole corresponds to infinitesimal translation
of the monopole,
and provides the functional measure \`{a} la collective coordinates. The grand canonical
partition function of the  monopole plasma is exactly equivalent to a local field 
theory with certain scalar fields interacting with the Yang-Mills fields. This integrates 
topological degrees of freedom with perturbation theory.
\end{abstract}
\newpage
\section{Introduction}
Consider the Yang-Mills quantum field theory in three Euclidean dimensions
given by the partition function
\bea \label{fi}
Z = \int D \A ~  \exp \left (-\frac{1}{2g^2} \int d^3x \B^a(x)\cdot \B^a(x) \right )
\eea 
where $\B^a(x)= \del \X \A^a (x) - \frac{1}{2} \epsilon^{abc} \A^b
(x) \X \A^c (x), a=1,2,3$, is the non-Abelian magnetic field in the vector 
representation of the group $SO(3)$.
The theory is super-renormalizable in a perturbation in the coupling
constant $g$. Therefore we have a good control on the ultraviolet properties
of the theory. The  perturbation theory has severe infrared divergences. 
It is expected that the theory is confining, so that there
is a mass gap and no infrared divergences. This is supported by
lattice gauge theory in a strong coupling expansion and also
simulations. We need a qualitative understanding of this phenomenon
and  quantitative techniques to handle it.

The Georgi-Glashow model in three Euclidean dimensions (GGM), which is a closely 
related model, is very well understood \cite{p}. (For a review see \cite{r1}.)
Fluctuations about a
(presumed) expectation value of the Higgs scalar suggest a massless
`photon', massive charged vector bosons and a massive Higgs. But
non-perturbative effects change the spectrum of the theory drastically.
The `dual photon' acquires a tiny mass and the theory is
confining. The reason is a `plasma' of monopoles and anti-monopoles which
screens the photon and forms a dipole sheet across a Wilson loop used as
a probe of confinement. The action has a stable, topologically non-trivial extremum --
the 't Hooft-Polyakov monopole \cite{thp} (playing the role of an instanton here). 
In a regime of coupling constants, a dilute gas approximation of
the plasma can be justified and the effects can be reliably computed.

The situation in case of pure gauge theory is much more complicated.
The  't Hooft-Polyakov ansatz for the Yang-Mills potential is
\bea \label{thp} 
A_i^a(x) = \epsilon _{iab} x^b \frac{1-K(r)}{r^2}
\eea
where $i=1,2,3$ labels the space index and $a,b=1,2,3$ label the group indices.  
Here $r=\sqrt{x^ax^a}$. As long as $K(r)=1+O(r^2)$ as $r\rightarrow 0$ and
$K(r)\rightarrow 0$ as $r\rightarrow \infty$, this configuration
is also of finite action in pure Yang-Mills theory.
But there is no non-trivial classically stable configurations with a finite
action. The configuration (\ref{thp}) is unstable against an indefinite expansion, as
may be checked by a rescaling. For $K(r)\rightarrow 1$, it is degenerate with the
perturbative vacuume $A =0$. We need to handle these long range fluctuations.
We need techniques very different from the case of GGM. Nevertheless we
may expect (Refs.\ \cite{r1}, \cite{bs}-\cite{r2}) that 
topological configurations are relevant for the infrared
behaviour of the theory.

The first question is whether the configurations (\ref{thp}) have any
topological significance. They have the telltale effects of a
monopole on a large Wilson loop. They contribute a phase proportional to
the solid angle subtended by the loop at the monopole (center). But we need
to charecterize how they are distinct from perturbative fluctuations.

For a configuration with many monopoles and anti-monopoles in GGM the
topological features can be located by the zeroes of the Higgs field and
its behaviour in the neighborhood \cite{afg, ap}. In Ref.\cite{ams,hms1} it has been proposed that
for the Yang-Mills field of (\ref{thp}), the topological aspects are charecterized by the
degeneracies in eigenvalues of the gauge invariant symmetric matrix
field, $ S_{ij}(x)=\sum_a B_i^a(x)  B_j^a(x) \, $, 
\bea  \label{ev}
S_{ij}(x) \xi^{jA}(x)=\lambda^A(x)  \xi^{iA}(x),~~ A=1,2,3
\eea 
The (anti-)monopoles can be located at points where the three eigenvalues become degenerate. 
At such points one of the eigenfunctions, say $ A=3 $, has a 'radial' behaviour.

We compare and contrast this proposal with the Abelian projection proposal of t'Hooft \cite{ap}. 
There a composite scalar transforming in the adjoint representation of the gauge group is 
formed out of the gauge field and used in place of
the fundamental Higgs of the GGM. The location of the zeroes depend on
the composite chosen, though the net `monopole charge' is an invariant.
In contrast we are using a gauge invariant composite  (\ref{ev}) for locating the topological
aspects. In spite of this difference there is a direct connection with the abelian projection
proposal. We also have a closely related entity, $ s^{ab}(x)=\B^a(x) \cdot \B^b(x) \, $, 
\bea  \label{ev1}
\B^a(x) \cdot \B^b(x) \xi_b^A=\lambda^A(x) \ \xi_a^A(x)
\eea
which transforms homogeneously in the symmetric tensor representation of gauge group  $SO(3)$.
Its eigenvalues are the same as of \ref{ev} and hence gauge invariant.
Each of the three eigenfunctions are in the adjoint representation. One of these, say $A=3$,
plays the role of the scalar composite of t'Hooft.

We have located the monopoles at points where the eigenvalues in (\ref{ev}) 
are triply degenerate. 
We may expect that a more generic singularity is a double
degeneracy as analysed by t'Hooft in the context of abelian projection \cite{ap}.  
In Ref. \cite{hms2} it has been argued that the generic configuration has
half-monopoles joined by (presumably) $Z_2$ strings. The triple degeneracy
is the extreme limit where the two half monopoles are collapsed on each other.
(It is likely that our criterion \cite{ams, hms1} gives a complete charecterization of the
topological singularities of the Yang-Mills fields in 3-dimensions.)
In this paper we restrict only to the point singularities i.e. the (anti-)monopoles.
It is possible that very long  $Z_2$ strings are the dominant configurations.
This  more general case can also be handled by our techniques.
This will be illustrated in case of Yang-Mills theory in 3+1 dimensions elsewhere \cite{hss}.

Our choice for locating the topological aspects has an added advantage.
As $s^{ab}(x)$ is a symmetric (real) matrix, the three eigenfunctions
$\xi_a^A(x)$  in (\ref{ev1}) (after normalization) form an orthonormal set and give
an SO(3) matrix which can be used for a
local gauge transformation. Note that a given Yang-Mills field uniquely
prescribes this gauge transformation (up to a trivial ambiguity in
the choice of the eigenvectors).

At points of degeneracy of the eigenvalues the gauge transformation is
singular. In case of the 't Hooft-Polyakov ansatz (\ref{thp}), the $SO(3)$ matrix takes 
the form $(\thetau, \phiu, \ru)$ i.e. the unit vectors of the spherical 
coordinate system \cite{thp, afg, hms1}. Even though the
Yang-Mills configuration (\ref{thp}) looks innocuous  the transformed gauge
field has singularities. Indeed, the third component (in group space) 
is precisely the Dirac vector
potential of a monopole. The Dirac string along the third direction is due
to an arbitrary choice of the `curling' eigenvectors $\thetau, \phiu $ among 
the degenerate eigenvectors. Thus our gauge transformation which is
dictated by the gauge configuration itself, highlights the topological aspects
of the given gauge field.

The value of a singular gauge configuration has been realized early
in attempts (Refs.\ \cite{bs}-\cite{r2}) to handle topological configurations beyond a
semiclassical approximation. There is extensive work using the maximal abelian
gauge. Our method of using the eigenfunctions of Eq.\ (\ref{ev1}) has many technical 
advantages as seen below.

Even though our gauge transformation has singularities, it is an $SO(3)$
matrix at each $x$. Therefore, even though the transformed gauge potential has
singularities, the tranformed non-abelian field strength, which was
finite to begin with, remains finite everywhere. This is the reason
why the Dirac string does not contribute a singular term to the action.
This is shown explicitly in Ref.\ \cite{hms3}. 

We have used the singular gauge transformation to motivate
the action in the
form to expose the topological degrees of freedom \cite{hms3}.
With this choice of gauge,
the transformed non-abelian magnetic fields $\B^1, \B^2, \B^3$
are mutually orthogonal.
However, we do not want to be
bound to this gauge except close to the locations of monopoles (where
we want all Dirac monopoles to be in the color diection $A=3$ and all
Dirac strings to be along the z-axis). We
can therefore continue using the Faddeev-Popov technique (eliminating the gauge zero-modes 
of the fields by gauge-fixing) and  keep the successes of
renormalized perturbation theory. 

We have abelianized the topological
aspects by having all monopoles in one colour direction. Also,
we have a linear superposition of the topological configurations, to
make any computation possible. Thus our action to include the topological
degrees explicitly is \cite{hms3}
\bea \label{ac}
S = \int d^{3}x  \left ( \frac{1}{2} (-\del \Phi
+ \nabla \X \abf + i \W^{+}  \X \W^{-} )^2
+ | \D [\A + \abf ] \X  \W^{-} |^2 \right )
\eea
where we have ignored the Dirac string contribution as justified in Ref. \cite{hms3}. Here
\bea
\Phi(x)=\sum_m q_m \frac{1} {| \x - \x_m|}
\eea
is the magnetic potential due to monopoles and anti-monopoles at $\x_m$ of (quantized) charges
$q_m $. Also
\bea \label{cd}
\D [\A]= \del -i\A
\eea
is the Abelian covariant derivative. In addition there are the gauge fixing and 
ghost terms which will not be handled explicitly in this paper.
 
Thus we have Dirac monopoles at arbitrary points with `photons' and charged 
massless vector bosons $\W^{\pm}$ scattering off them. Even though the monopole field
is singular, the action is rendered finite by the singular boundary conditions
of $\W$ at the location of the monopoles. In fact requiring the action be finite
selects out the boundary conditions for the charged vector mesons.
\bea
\W(x) \rightarrow (\phiu_m - i \thetau_m)\frac{e^{i\phi_m}}{\sqrt 2 |\x-\x_m|} 
+ O(| \x - \x_m|).
\eea
where $\phi_m, \theta_m$ are the spherical coordinates centered at $\x_m$.

We now point out how the topological aspects change the contribution to
the partition function. In the action (\ref{ac}) consider terms quadratic in $\W$, 
setting the 'photon'
fluctuations $\abf$ to zero. This has precisely the form $\W^{+} H \W^{-} \, $ where H is the
non-relativistic Hamiltonian (with unit mass) for a charged vector boson
interacting with Dirac monopoles at fixed positions \cite{hms3}.(This includes an
anomalous magnetic moment $g=2$ interaction, which is a consequence of non-abelian
gauge invariance). This hamiltonian has zero modes \cite{hms3} of the form $\D \Lambda$ for an
arbitrary function $\Lambda(x)$ where $\D$ is the covariant derivative \ref{cd}. 
This is just a reflection of
the  local gauge invariance. With any gauge fixing for $\W$, these zero modes are absent.
In case of one monopole, it has been shown in Ref\cite{hms3} that
there are no other zero modes. Thus (after a gauge fixing) we just get
a contribution $({\rm Det} H)^{-1}$. This determinant involving  the background of the
Dirac monopoles depends on the scattering phase shifts and is obviously different
from that of free vector bosons. In this way the monopole configurations of
the Yang-Mills theory give a contribution distinct from the perturbation theory.

\section{Local field theory of interactions of monopole charges and electric
currents in three Euclidean dimensions} \label{dual}

Consider the free Maxwell theory in three Euclidean dimensions.The partition function is
\bea \label{m}
Z = \int D \abf ~  \exp \left (-\frac{1}{2g^2} \int d^3x (\del \X \abf (x))^2 \right )
\eea
Using an auxiliary field $\bbf$ we write
\bea \label{m1}
Z = \int D \bbf  D \abf ~  \exp \left (\int d^3x (-\frac{g^2}{2} \bbf(x)^2
+i\bbf(x) \cdot \nabla \X \abf (x)) \right )
\eea
Note the presence of $i=\sqrt{-1}$ in the term linear in $\abf$ in the exponent.
Now the integration over $\abf$  gives a constraint:
\bea
Z = \int D \bbf   ~ \prod_x \delta(\nabla \X \bbf(x)) \exp \left (-\frac{g^2}{2} \int d^3x \bbf(x)^2 \right )
\eea
The  constraint has the solution $\bbf(x)=\del \chi(x)$. Therefore
\bea
Z = \int D \chi ~  \exp \left (-\frac{g^2}{2} \int d^3x (\del \chi (x))^2 \right )
\eea
describing a free massless scalar. This field has the interpretation as the `dual' photon.
In three dimensions the `photon' has only one transverse degree of freedom
which is described by the scalar $\chi$.

Consider now the interaction between sources $j$ of Dirac monopoles and
electric currents $\J$. This is described by the partition function
\bea \label{me}
Z&=&\exp\Bigg(\int d^3x d^3y\Big(-\frac{1}{2g^2}j(x)D(x-y)j(y)+i\, j(x)A_i(x-y)J_i(y)
                                                \nonumber\\
              && -\frac{g^2}{2}J_i(x)D_{ij}(x-y)J_j(y)\Big)\Bigg)
\eea
$D(x-y)$ and $D_{ij}(x-y)$
are the free propagators for massless scalar (monopole)
 and massless vector fields in coordinate space.
$A_i(x-y)$ is the Dirac vector potential of a monopole, here playing the role of the
propagator connecting monopole charges to electric currents. Note the factor $\sqrt{-1}$ in 
the interaction of magnetic monopoles and electric currents, and also the coupling constants
appearing in inverse proportions in (\ref{me}). 

The partition function in (\ref{me}) describes the following local quantum field 
theory involving monopole charges, photons and electric currents:
\bea \label{me'}
Z &=& \int D\abf D \chi ~  \exp \Bigg(\int d^3x \Big(-\frac{1}{2} (\partial_3 \chi (x))^2
-\frac{1}{2}(\nn_{3} \X (\del \X \abf(x)))^2  \nonumber\\
&&+i \partial_3 \chi(x) \nn_{3} \cdot \del \X \abf(x)+ ig J_i(x)a_i(x)+ig^{-1} j(x) \chi (x) 
\Big)\Bigg)
\eea
To see this, we use a gauge-fixing term for $\abf(x)$ corresponding to the Feynman gauge
and go over to the Fourier space. [It is useful to employ the identity
\bea
(\nn_{3} \X (\del \X \abf))^2=(\del \X \abf)^2-(\nn_{3}\cdot \del \X \abf)^2
=(\del \X \abf)^2-((\nn_{3} \X \del)\cdot\abf)^2
\eea
to simplify (\ref{me'}).] Then we obtain
\bea
Z &=& \int D\abf(k) D \chi(k) ~  \exp \Bigg(\int d^3k \Big(-\frac{1}{2}\chi(-k){k_3}^2\chi(k)
-\frac{1}{2} a_i(-k)(k^2 \delta_{ij}-{k_\perp}_i{k_\perp}_j)a_j(k)          \nonumber\\
&&+i\chi(-k)k_3{k_\perp}_i a_i(k)
+ig J_i(-k)a_i(k)+ig^{-1} j(-k) \chi (k)
\Big)\Bigg)
\eea
where $\kv_\perp=\nn_{3} \X\kv$. On performing the functional integration over
$\abf(k)$ and $\chi(k)$, one gets a quadratic form in the currents. This is found
to be the same as that in Eq.\ (\ref{me}) in Fourier space, on using the momentum space
representation of $A_i(x-y)$ given in Eq.\ (\ref{Amom}).

Eq.\ (\ref{me'}) has the `photon' quanta described simultaneously by the dual fields, 
the usual vector potential $\abf$
and the dual scalar $\chi$. This is the analogue of the two potential formalism of Zwanziger 
\cite{zw}
for quantum electrodynamics of monopoles and charges in three Euclidean dimensions. 
Note the characteristic
appearance of $\nn_{3} \cdot \del \X \A$  in the terms involving the photon and the dual photon.
Also only the $\partial_3 $ derivative of $\chi$ is appearing. Similar terms will
appear in Sec.\ \ref{plasma}.

\section{Functional measure in the presence of monopole configurations}

We now address the issue of the functional measure in the presence of
monopoles. In the case of a semiclassical quantization about (say) an instanton,
the standard technique to obtain the measure is the collective coordinate method.
The quadratic terms in fluctuations about an instanton have zero modes
related to translation and other continuous symmetries of the theory
that are broken by the choice of position and other degrees of freedom of the instanton.
The fluctuations which translate the instanton (for example)
are replaced by an integration over the position of the instanton using the Faddeev-Popov trick.

We now point out that even though we are not doing an expansion about a saddle
point of the action, we can adopt the same strategy.
The action in Eq.\ (\ref{ac}) contains the combination $-\del \Phi
+ \nabla \X \abf$. 
Given a magnetic potential $\Phi$ corresponding to a configuration of monopoles and
anti-monopoles of various (quantized) charges, consider another where all
but one (anti-)monopole, say at $x_m$, is displaced by an infinitesimal amount
$\delta_j$ in the $j$th direction. The difference in the  magnetic potential $\Phi$ of the
two configurations is precisely that of a magnetic dipole of moment
$q_m \delta_j$. The corresponding magnetic field $-q_m \delta_j \partial_j \del \phi$ can be
obtained from a vector potential. 
This means that the mode of the `photon' $\abf$ corresponding to such a dipole can be treated 
as a mode which displaces the monopole at $x_m$ in the $j$th direction. 
We can therefore eliminate theses modes form  $\abf$ and replace them with integration over
the positions of the monopoles. 

Let us now directly check that these dipole modes of  $\abf$ are of the form
$\partial_j \A(x-x_m)$. The Dirac potential of a monopole located at the origin,
with the Dirac string along the $z$-direction, is given by Eq.\ (\ref{Dp}).  
First consider the derivative in the $z$-direction:
\bea
\frac{\partial \A(x)}{\partial x_3} = \frac{\nn_3 \X \rv}{r^3}
\eea
[This follows from Eq.\ (\ref{Ac}), Eq.\ (\ref{Ag}) and  $\partial_3 g=1/r$.]
This is precisely the vector potential of a dipole at the origin, and so the curl of it
gives the magnetic field
of a unit dipole moment in the z-direction.
But the situation is a little more involved for the derivatives along (say) the $x$
direction. The reason is that the Dirac string along the $z$
direction is now infinitesimally displaced in the transverse direction and its
effects persist in the vector potential. Consider the vector potential of a monopole
with the Dirac string in the $x$ direction. It differs from (\ref{Dp}) by a gauge
transformation. Its derivative in the $x$ direction gives the standard dipole
potential. Thus the $x$ and $y$ derivatives of (\ref{Dp})  also give the dipole potential
but in a non-standard gauge.

We insert a unit factor in the partition function for each (anti-) monopole:
\bea \label{cc}
1=\int d^{3} x_m \prod_{j=1,2,3} \delta ( \int d^{3}x \partial_j \A(x-x_m) \cdot \abf(x))
|\int d^{3}x \partial_p \partial_q \A(x-x_m) \cdot \abf(x)|
\eea
Here $\partial_p$ stands for derivative with respect to the $p$th component of
$ x_m$ and $|M_{pq}|$ stands for the determinant of a $3\times 3$ matrix $M_{pq}$.
The constraint is independent of the charge of the monopole. We  can use the BRST
techniques to handle these constraints. We have thus obtained integration over the
positions of the monopoles from the functional measure, 
\`{a} la the collective coordinate method.

\section{Effects of the monopole plasma from a local action}\label{plasma}
A major reason for the success \cite{p} in understanding the GGM is that the grand canonical
partion function of the monopole plasma could be handled. We need to know the
effects of an arbitrary number of monopoles and anti-monopoles at arbitrary locations on an
external probe. In order to do this in the present case we use the first order
formalism. Note that the first order formalism is as good as the usual second order formalism for 
carrying out a renormalized perturbation theory.
For explicit Feynman rules and diagrammatic calculations (in the 3+1-dimensional context) 
see Ref.\ \cite{z}.

In the first order formalism, our  partition function takes the form (as in Eq.\ (\ref{m1}))
\bea \label{fo}
Z&=& \int D \abf D \W^{-} D \W^{+} D\bbf  D\bbf^{+} D\bbf^{-}~  \exp \Bigg(\int d^{3}x   
\Big(-g^2 (\frac{1}{2}\bbf^2+\bbf^{+} \cdot \bbf^{-})
+i  \bbf \cdot (-\del \Phi \nonumber\\ 
&&+ \del \X \abf 
+ i \W^{+}  \X \W^{-} ) 
+ i\bbf^{+} \cdot \D [\A + \abf ] \X  \W^{-} 
+i\bbf^{-} \cdot \D [\A + \abf ] \X  \W^{+}\Big) \Bigg)
\eea
There are also the gauge fixing and the ghost terms, which we have not written explicitly.
Consider the gradient and curl parts of  $ \bbf$:
\bea \label{gc}
\bbf=\del \chi+\del \X \cbf
\eea
Now the part of the partition function involving $\bbf$ takes the form,
\bea \label{fo1}
Z&=& \int D \abf D \W^{-} D \W^{+}  D\bbf^{+} D\bbf^{-} D\chi D\cbf
~  \exp \Bigg(\int d^{3}x\Big(-\frac{g^2}{2}((\del \chi)^2+(\del \X \cbf)^2) \nonumber\\
&&+ \chi(x) \del \cdot (\W^{+}  \X \W^{-})+i(\del \X \cbf )\cdot (\del \X \abf + i \W^{+}  \X \W^{-}) 
-i\sum_{m}q_m \chi(x_m) \nonumber\\
&&+\rm{other~ terms}\Big) \Bigg)
\eea
Here we have used
\bea
\del^2 \Phi(x)=-\sum_{m}q_m \delta(x-x_m)
\eea
The gradient part $ \chi$ is the `dual photon' as discussed in Sec.\ \ref{dual}. 
In Eq.\ (\ref{fo1}), $\chi$  couples to the monopoles 
locally as the dual photon potential should.
If we consider only monopoles and anti-monopoles of unit charge as in \cite{p}, 
summing over these charges we get 
$\cos \chi(x)$. A sum over arbitrary number of monopoles and anti-monopoles
exponentiates this  into a new term in the action \cite{p}. This gives
a mass to the `dual photon'. However the situation is more complicated in our case.

We also have to handle the interactions of the monopoles with the massless charged vector bosons.
It is possible to adopt the techniques of Sec.\ \ref{dual} with some
modification to obtain the analogue of the two potential formulation. But
we adopt a different procedure here. Our main purpose here is to represent
net effects of the monopole plasma by a local field theory. 

The interaction of a monopole at $x_m$ with an electric current at $x$ can be 
put in the convenient form
\bea\label{conv}
i \int d^{3}x \A(x-x_m)\cdot \J(x)=i\,4\pi\int d^{3}x (\partial_3 \nabla^2) ^{-1}(x-x_m)
\nn_{3} \cdot \del \X \J(x)
\eea
where we used Eq.\ (\ref{Axx'}). Here the specific form of the (hermitean) current $\J$ 
will not matter. [In Eq.\ (\ref{fo}),
$\J=i(\bbf^{+}\X  \W^{-}
-\bbf^{-} \X  \W^{+})$.] Now, Eq.\ (\ref{conv}) will have a sum over $m$ corresponding to 
monopole charges $q_m$ in the present case. The resulting expression can be put in a local form
using auxiliary scalars $\phi$ and $\psi$:
\bea \label{vb}
\int D \phi D \psi ~  \exp \Bigg(i\int d^{3}x \psi(x)\partial_3 \nabla^2 \phi(x) 
+i\sum_{m}q_m \phi(x_m)+i\, 4\pi\int d^{3}x\psi(x)\nn_{3} \cdot \del \X \J(x) \Bigg)
\eea
[To see this, write the second term in (\ref{vb}) as 
$i\sum_{m}q_m\int d^{3}x\,\phi(x)\delta(x-x_m)$. Integration over $\phi(x)$ then gives
a constraint on $\psi(x)$.] 
Thus the scalar $\phi$ couples locally to the monopoles just as $\chi$ in Eq. (\ref{fo1}).

Now we consider the effects of summing over the monopole charges $q_m=\pm 1, \pm 2$ etc. 
In GGM \cite{p}, the fugacity for each monopole meant that it is sufficient to
consider only $q_m=\pm 1$. Now there is no seperate fugacity for each monopole. Nonetheless 
we expect the generic configuration will have $q_m=\pm 1$, with higher charges resulting
from a merging of monopoles with a loss of entropy. Summing over only $q_m=\pm 1$ we get
$\cos (\phi(x)-\chi(x))$ in the action. The combination $\phi(x)-\chi(x)$ plays the role of 
$\phi(x)$ of the GGM.

We now address the constraints due to the collective coordinates (\ref{cc}). These are
non-local constraints on the `photon'. We make them local by introducing an auxiliary
scalar $\eta(x)$:
\bea
\sum_{N=0}^{\infty} \frac{1}{N!} \int D \eta \prod_{x} 
\delta(\partial_3 \del^2 \eta(x) +4\pi \nn_{3} \cdot \del \X \abf(x))\nonumber\\
\times\prod_{m=1}^N \int d^{3} x_m \prod_{j=1,2,3} \delta ( \partial_j \eta(x_m))
|\partial_p \partial_q \eta(x_m)|                                    \label{eta}
\eea
[To check this, define $\eta(x)=\int d^{3}y \A(y-x) \cdot \abf(y)$,
use Eq. (\ref{Axx'}) and operate on $\eta(x)$ by $\partial_3 \del^2$.]
Eq.\ (\ref{eta}) means that the auxiliary scalar $\eta$ has an extremum at the locations of the
(anti-)monopoles, and $ |\partial_p \partial_q \eta|$ is the determinant of the
quadratic fluctuations at these extrema.

For the collective coordinates we can as well require the `photon' fluctuations to satisfy 
the constraint $\int d^{3}x \partial_j \A(x_m-x) \cdot \abf(x)= \vbf_j$ for any chosen vector $\vbf$.
This allows a smoothening of the constraint $\delta ( \partial_j \eta(x_m))$ as in the
Faddeev-Popov procedure. Also for the determinant we can use ghost fields. 

In this paper we developed a  procedure to rewrite the Yang-Mills quantum field theory
in three Euclidean dimensions to formally include the monopole configurations. The effects 
can be presented as a local action with some new scalar fields. The connection to the two
-potential formulation of Zwanziger \cite{zw} is made. Our aim is to obtain a reliable scheme 
of calculations which integrates renormalised perturbation theory for the ultraviolet behaviour 
with effects of topological degrees for the infrared behaviour.

\section*{Acknowledgement}
I.M. thanks IMSc, Chennai for hospitality during the
course of this work.

\appendix
\leftline{\null\hrulefill\null}\nopagebreak
\section{Appendix}
In this Appendix, we represent the Dirac vector potential of a monopole in a more amenable form.
The Dirac potential of a monopole located at the origin has the form
\bea
\vec A(x)=\phiu\frac{\sin\theta}{r(1+\cos\theta)}=\nn_3\X\frac{\ru}{r+x_3}     \label{Dp}
\eea
with the Dirac string along the negative $z$-direction.
Here $\nn_3$ is the unit vector along $z$-direction.
[For checking this and other
results below, a useful formula is
$\nn_3=\cos\theta\ru-\sin\theta\thetau$.]
Let us write
\bea
\A(x)=\nn_{3} \X \cbf                                                \label{Ac}
\eea
where the vector field $\cbf$ is undetermined upto
addition of a vector in the 3-direction. We choose
\bea
\cbf=\frac{\ru+\nn_3}{r+x_3}
\eea
so that \cite{baal}
\bea
\cbf = \del g, ~~~g=\ln(r+x_3)                                       \label{Ag}
\eea
Note that $\partial_3 g=1/r$, and so $\partial_3 \del^2 g=-4\pi\delta(x)$.
Thus the Dirac potential at $x$ due to a monopole at $x'$ can be expressed in terms of
the Green function for the operator $\partial_3 \del^2$:
\bea
\A(x-x')&=&\nn_3\X\del\ln(|x-x'|+x_3-x'_3)            \\
        &=&-4\pi \nn_3\X\del (\partial_3 \nabla^2) ^{-1}(x-x')    \label{Axx'}
\eea
[An alternative form of the Dirac potential is $\A=-\phiu(1/r)\cot\theta
=-\nn_3\X\ru (x_3/\rho^2)$, with the Dirac strings along the $\pm z$ directions.
Here $\rho^2={x_1}^2+{x_2}^2$.
In this case, we choose $\cbf=(r\nn_3-x_3\ru)/\rho^2$ in Eq.\ (\ref{Ac}).
Then
$\cbf = \del g$ and  $\partial_3 g=1/r$ continue to hold, but  
with $g=(1/2)\ln((r+x_3)/(r-x_3))$.]

The result given in Eq.\ (\ref{Axx'}) can
also be seen by going over to the Fourier space. For the potential of Eq.\ (\ref{Dp}),
\bea
\del \X \A= \frac{\ru}{r^2}+\nn_{3}4\pi\delta(x_1) \delta(x_2)\theta(-x_3)
\eea
Taking the Fourier transform, we get
\bea
\kv \X \At(\kv)=-4\pi\frac{\kv}{k^2}+\frac{4\pi}{k_3 }\nn_{3}
\eea
(The Fourier transform of the theta function can be obtained using
$d\theta(x)/dx=\delta(x)$.) We now evaluate $\kv \X$ both sides and use
$\kv\cdot\At=0$ (since $\del\cdot\A=0$) to obtain
\bea
\At(\kv)=\frac{4\pi \nn_{3}\X\kv}{k_3 k^2}                  \label{Amom}
\eea
This agrees with Eq.\ (\ref{Axx'}).

\end{document}